\documentclass[10pt]{IEEEtran}

\usepackage[linesnumbered,boxed]{algorithm2e}
\usepackage{epstopdf}
\usepackage{graphicx}
\usepackage{stmaryrd}
\usepackage{amssymb,amsmath}
\usepackage{multirow,url}
\usepackage{color,cite}
\usepackage{slashbox}

\newtheorem{lemma}{Lemma}
\newtheorem{corollary}{Corollary}
\newtheorem{proposition}{Proposition}
\newtheorem{definition}{Definition}
\newtheorem{theorem}{Theorem}
\newtheorem{observation}{Observation}

\ifodd 0
\newcommand{\rev}[1]{{\color{blue}#1}} 
\newcommand{\revh}[1]{{\color{magenta}#1}} 
\newcommand{\com}[1]{\textbf{\color{red} (COMMENT: #1) }} 
\newcommand{\comg}[1]{\textbf{\color{green} (COMMENT: #1)}}
\newcommand{\response}[1]{\textbf{\color{green} (RESPONSE: #1)}} 
\else
\newcommand{\rev}[1]{#1}
\newcommand{\revh}[1]{#1}
\newcommand{\com}[1]{}
\newcommand{\comg}[1]{}
\newcommand{\response}[1]{}
\fi

\begin{document}

\title{Contract-Based Cooperative Spectrum Sharing}

\author{Lingjie Duan, Lin Gao, Jianwei Huang\\
Department of Information Engineering, the Chinese University of
Hong Kong, Hong Kong\\
email: \{dlj008, lgao, jwhuang\}@ie.cuhk.edu.hk}

\maketitle


\begin{abstract}

Providing proper economic incentives is essential for the success of
dynamic spectrum sharing. \emph{Cooperative spectrum sharing} is one
effective way to achieve this goal. In cooperative spectrum sharing,
secondary users (SUs) relay traffics for primary users (PUs), in
exchange for dedicated transmission time for the SUs' own
communication needs. In this paper, we study the cooperative
spectrum sharing under \emph{incomplete information}, where SUs'
types \rev{(capturing their heterogeneity in relay channel gains and
evaluations of power consumptions) }are private information and not
known by PUs. Inspired by the contract theory, we model the network
as a labor market. The single PU is the employer who offers a
\emph{contract} to the SUs. The contract consists of a set of
contract items representing combinations of spectrum accessing time
(i.e., reward) and relaying power (i.e., contribution). The SUs are
employees, and each of them selects the best contract item to
maximize his payoff. We study the optimal contract design for both
weak and strong incomplete information scenarios. First, we provide
necessary and sufficient conditions for feasible contracts in both
scenarios. In the weak incomplete information scenario, we further
derive the optimal contract that achieves the same maximum PU's
utility as in the complete information benchmark. In the strong
incomplete information scenario, we propose a Decompose-and-Compare
algorithm that achieves a close-to-optimal contract. We  future show
that the PU's average utility loss due to the suboptimal algorithm
and the strong incomplete information are both relatively small
(less than $2\%$ and $1.3\%$, respectively, in our numerical results
with two SU types).
\end{abstract}




\section{Introduction}

{With the explosive development of wireless services and networks,
spectrum is becoming more congested and scarce. }Dynamic spectrum
sharing is a promising approach to increase  spectrum efficiency and
alleviate spectrum scarcity, as it enables the unlicensed secondary
users (SUs) to dynamically access the spectrum licensed to the
primary users (PUs) \cite{Akyildiz-survey, Marshall1, buddhikot-sur,
Mahanen}. The successful implementation of dynamic spectrum sharing
requires many innovations in technology, economics, and policy. In
particular, it is important to design the sharing mechanism such
that PUs have incentive to open their licensed spectrum for sharing,
and SUs have incentive to utilize the new spectrum opportunities
despite of the potential costs.

\begin{figure*}[tt]
\centering
\includegraphics[width=0.9\textwidth]{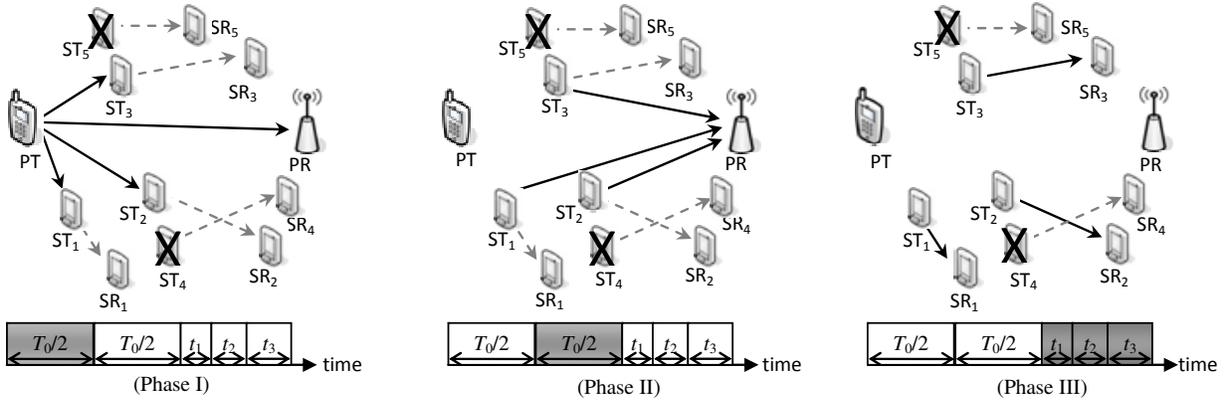}
\caption{Cooperative spectrum sharing with three phases in each time
slot} \label{fig:network}
\end{figure*}

Market-driven {spectrum trading} is a promising paradigm to address
the incentive issue in dynamic spectrum sharing. With spectrum
trading, PUs temporarily \emph{sell} the spectrum to SUs to obtain
either a monetary reward or a performance improvement. A particular
interesting trading scheme is \emph{cooperative spectrum sharing},
where SUs relay traffics for PUs in order to get their own share of
spectrum. A brief illustration of cooperative spectrum sharing is
shown in Figure \ref{fig:network} \rev{on the next page}. The SUs'
transmitters  (ST$_1$ $\sim$ ST$_3$) act as cooperative relays for
the PU in Phase I and Phase II (Decoding and Forwarding), and
transmit their own data in Phase III.

Researchers have only recently started to study cooperative spectrum
sharing mechanisms \cite{cocr1,cocr2,cocr3,cocr4}. The prior results
all assumed complete network information, i.e., PUs know SUs'
channel conditions, resource constraints, and costs of transmission.
This assumption is often too strong for practical networks.
In this paper, we study the cooperative spectrum sharing  under
\emph{incomplete information}.
\rev{We consider the general case that SUs have different
\emph{types} based on their relay channel gains and evaluations of
power consumptions.} The types are private information, and only an
SU knows his own type.

To tackle this problem, we propose a contract-based cooperative
spectrum sharing mechanism. Contract theory is an effective tool in
designing the incentive compatible mechanism in a monopoly market
under incomplete information\cite{ContractTheory}. \rev{The key idea
is to offer the right contract items so that all of the agents have
the incentive to truthfully reveal their private information.} For
the spectrum sharing problem, we can imagine the network as a labor
market. The PU is the employer and offers a \emph{contract} to the
SUs. The contact consists of a set of contract items, which are
combinations of spectrum access time \rev{(i.e., reward) }and relay
power \rev{(i.e., contribution)}. The SUs are employees, and each SU
selects the best contract item according to his type. \emph{We want
to design an optimal contract that maximizes the PU's utility
(average data rate) under the incomplete information of SUs' types}.

The main contributions of this paper are as follows:

\begin{itemize}

\item
\emph{New modeling and solution technique}: As far as we know, this
is the first paper that tackles cooperative spectrum sharing under
incomplete information based on contract theory.

\item
\emph{Multiple information scenarios}: We study the optimal contract
designing in three different scenarios: complete information
(benchmark), weak incomplete information, and strong incomplete
information. In the last two incomplete information scenarios, the
PU does \emph{not} know the exact type of each SU. The difference
between the two scenarios is whether the PU knows the number of SUs
in each type (the weak scenario) or only knows the distribution of
types (the strong scenario). We will design optimal contracts for
all three scenarios.

\item
\emph{Sufficient and necessary condition for feasible contracts}:
Under incomplete information, a contract is feasible if and only if
it satisfies the incentive-compatibility (IC) and individually
rationality (IR) for each SU. We find necessary and sufficient
conditions that satisfy both IC and IR conditions. \rev{This
discussion helps us understand important properties of a contract
and how to design optimal contracts later.}

\item
\emph{Optimal contract design}: In the weak incomplete information
scenario, we derive the optimal contract that achieves the same
maximum PU's utility as in the complete information benchmark. In
the strong incomplete information scenario, we propose a
Decompose-and-Compare algorithm that obtains a close-to-optimal
contract.

\item
\emph{Performance analysis}: In the strong incomplete information
scenario, we quantify the PU's average utility loss due to the
suboptimal algorithm (by comparing it with the exhaustive search
method) as well as the strong incomplete information (by comparing
it with the complete information benchmark). Both kinds of losses
are are relatively small, i.e., less than $2\%$ and $1.3\%$ in our
numerical results with two SU types, respectively.
\end{itemize}

The key results and the corresponding section numbers in this paper are are summarized in Table \ref{major-work}.

\begin{table}[t]
\renewcommand{\arraystretch}{1.5}
\centering
\caption{Feasibility conditions and Optimality}
\label{major-work}
\begin{tabular}{|c|c|c|c|}
\hline \textbf{Network Information}  & \textbf{Feasibility} &
\textbf{Optimality}  & {\renewcommand{\arraystretch}{1}
\hspace{-5mm}
\begin{tabular}{c}
                            \textbf{Sections in} \\
                            \textbf{this paper}
                            \end{tabular}
\hspace{-5mm}
}
\\
\hline
 \textbf{Complete (benchmark)} & IR & Optimal & \ref{sec:complete} \\
\hline
 \textbf{Weak Incomplete} & IC \& IR & Optimal & \ref{sec:feasible}, \ref{sec:incomplete1} \\
\hline
 \textbf{Strong Incomplete} & IC \& IR & Close-to-Optimal & \ref{sec:feasible}, \ref{sec:incomplete2}\\
\hline
\end{tabular}
\end{table}

The rest of this paper is organized as follows. In Section
\ref{sec:system}, we provide the system model and problem
formulation. In Section \ref{sec:complete}, we propose the optimal
contract under complete information (benchmark). In Section
\ref{sec:feasible}, we propose the necessary and sufficient
conditions for feasible contracts under incomplete information. In
Sections \ref{sec:incomplete1} and \ref{sec:incomplete2}, we derive
the optimal and suboptimal contracts in weak and strong incomplete
information scenarios, respectively. We present the numerical
results  in Section \ref{sec:Numerical}. We review the related
literatures in Section \ref{sec:RelatedWork} and finally conclude in
Section \ref{sec:Conclusion}.


\section{System Model and Problem Formulation}\label{sec:system}

We consider a cognitive radio network with a primary licensed user
(PU) and multiple secondary unlicensed users (SUs) as shown in
Fig.~\ref{fig:network}. Each user is a dedicated
transmitter-receiver pair. The PU has the exclusive usage right of
the licensed spectrum band, but its transmission suffers from the
poor channel condition between its transmitter PT and receiver PR.
We represent $M$ SUs by distinct transmitter-receiver pairs
$\{ST_k-SR_k\}_{k=1}^{M}$. Each SU wants to have dedicated time to
access the licensed band and transmits its own data. The PU can
employ a subset of or all SUs to relay its traffic; the
\emph{involved} SUs will obtain dedicated transmission time for
their own data. The interaction between the PU and the SUs involves
three phases as in Fig.~\ref{fig:network}: Phases I and II for the
cooperative communications with a total fixed length of time
$T_0$,\footnote{The time period $T_0$ is a constant and is
determined by PU's MAC layer and Physical layer specifications.} and
Phase III for the SUs' own transmissions. More specifically, we have
\begin{itemize}
\item Phase I: In the first half of the cooperative communications ($T_0/2$),
primary transmitter PT broadcasts its data to primary receiver PR
and the involved SUs' transmitters (e.g., $ST_1$, $ST_2$ and $ST_3$
in Fig.~\ref{fig:network}). Note that SU $4$ and SU $5$ are not
involved in this example.
\item Phase II: In the remaining half of the cooperative communications ($T_0/2$),
the involved SUs' transmitters (STs) decode the data received in
Phase I and forward to PU's receiver PR simultaneously using the
space-time codes assigned by PU.\footnote{We will discuss the
details of the coordination between PU and SUs in
Section~\ref{subsec:SU}.} Through proper choice of space-time codes,
SUs' simultaneous relay signals do not interference with each other
at the primary receiver PR\cite{Laneman,luo2007decode}.
\item Phase III: PU rewards each involved SU with a
dedicated time allocation for that SU's own data (e.g.,
$\{t_k\}_{k=1}^3$ for three involved SUs). SUs access the spectrum
using TDMA and do not interfere with each other.
\end{itemize}

\com{This paragraph not clear. What to screen? How to screen? Even
if the SUs can successfully decode in Phase I, it does not mean that
the Phase II will be bottleneck (as Phase II channel can be even
better or transmission power is larger than Phase I)?} \rev{Here we
assume that each involved SU can successfully decode PU's data in
the first phase of cooperative communications. Thus we can focus on
the relay links between the STs and PR, which are the performance
bottleneck of cooperative communications. This assumption can be
relaxed if the PU can perform an initial screening over all SUs as
follows. The PU first broadcasts a pilot signal to all SUs, and only
those SUs replying correctly can choose to accept the contract and
involve in the cooperative communications later. Notice that the
involved SUs are often the ones that are close to the primary
transmitter PT, and thus are not close to the primary receiver PR.
This explains why a $PT-ST$ channel is often better than the
corresponding $ST-PR$ channel.}
%

The PU and SUs have conflicting objectives in the above
interactions. The PU wants the SUs to relay its traffic with high
power levels, which will increase the PU's data rate but reduce the
SUs' battery levels. An SU $k$ wants to obtain a large dedicated
transmission time ${t_k}$, which will increase the SU's own
performance but reduce the PU's utility (time average data rate). In
Sections~\ref{subsec:PU} and \ref{subsec:SU2}, we will explain in
details how the PU and SUs evaluate the trade-off between relay
powers and time allocations. In Section~\ref{subsec:SU}, we propose
a contract-based framework which brings the PU and SUs together and
resolve the conflicts.


\subsection{Primary User Model}\label{subsec:PU}
In this subsection, we discuss how PU evaluates relay powers and
time allocations.

We first derive the PU's achievable data rate during the cooperative
communications (i.e., Phases I and II in Fig.~\ref{fig:network}).
Let us denote the set of involved SUs as $\mathcal{N}$ (e.g.,
$\mathcal{N}=\{1,2,3\}$ in Fig.~\ref{fig:network}). The received
power (at the primary receiver PR) from SU $k$ is $p_k$, and the
time allocation to this SU is $t_k$. \emph{Without loss of
generality, we normalize $T_0$ to be 1 in the rest of the paper.
Then $t_k$ can be viewed as the ratio $t_k/T_0$}.
\begin{itemize}
\item In Phase I, PU's transmitter broadcasts its data, and PU's receiver achieves a data rate (per unit time) of
\begin{equation}R^{dir}=\log(1+\mathtt{SNR}_{PT,PR}).\end{equation}
$R^{dir}$ remains as a constant throughout the analysis.
\item  In Phase II, each involved SU successfully decodes PU's
data and forwards to PU's receiver.
\end{itemize}

Thus the PU's total transmission rate during the cooperative
communications (Phases I and II) is (\cite{Laneman})
\begin{align}\label{eq:PU_rate}
r_{PU}^{relay}&=\frac{R^{dir}}{2}+\frac{1}{2}\log\left(1+\sum_{k\in\mathcal{N}}\mathtt{SNR}_{ST_k,PR}\right)
\nonumber\\&=\frac{R^{dir}}{2}+\frac{1}{2}\log\left(1+\frac{\sum_{k\in\mathcal{N}}p_k}{n_0}\right),
\end{align}
where $n_0$ is the noise power, and constant ${1}/{2}$ is due to
equal partition of the cooperative communication time into Phase I
and Phase II. We can think (\ref{eq:PU_rate}) as the sum of
transmission rates of two ``parallel'' channels, one from the PT to
PR and the other from the set of involved SUs' transmitters to PR.

Based on the above discussion, we next compute the PU's average data
rate during entire time period (i.e., Phases I, II, and III). The
cooperative communications only utilizes
${1}/{(1+\sum_{k\in\mathcal{N}}t_k)}$ fraction of the entire time
period. The PU's objective is to maximize its \textbf{utility}
(i.e., average transmission rate during the entire time period) as
follows
\begin{align}\label{eq:PU_utility}
u_{PU}&=\frac{1}{1+\sum_{k\in\mathcal{N}}t_k}r_{PU}^{relay}\\
\nonumber&=\frac{1}{1+\sum_{k\in\mathcal{N}}t_k}\left(\frac{R^{dir}}{2}+\frac{1}{2}\log\left(1+\frac{\sum_{k\in\mathcal{N}}p_k}{n_0}\right)\right),
\end{align}
which is decreasing in total time allocations to SUs
($\sum_{k\in\mathcal{N}}t_k$), and is increasing in the total
received power from SUs ($\sum_{k\in\mathcal{N}}p_k$).


We want to emphasize that the utility calculation in (\ref{eq:PU_utility})
assumes that the PU involves at least one SU in the cooperative
communication. The PU can also choose to have direct transmissions only
in both Phases I and II and does not interact with the SUs (and thus
there is no Phase III). The total rate in this direct transmission
only approach is $R^{dir}$. This means that the PU will only choose
to use cooperative communications if the utility in (\ref{eq:PU_utility}) is larger than $R^{dir}$. In the rest of
the analysis, we assume that $R^{dir}$ is
small such that the PU wants to use cooperative communications. In
Section \ref{sec:Numerical}, we will further explain what will happen when this is not true.

\subsection{Secondary User Model}\label{subsec:SU2}
Next we discuss how SUs evaluate relay powers
$\{p_k\}_{k\in\mathcal{N}}$ and time allocations
$\{t_k\}_{k\in\mathcal{N}}$. \emph{We want to emphasize that the
relay power is measured at the PU's receiver, not at the SUs'
transmitters.} We consider a general model where SUs are
heterogeneous in three aspects:
\begin{itemize}
\item SUs have different relay channel gains between their transmitters
and the PU's receiver ($h_{{ST},PR}$). If an SU $k$ wants to reach a
received power $p_k$ at the PU's receiver, then it needs to transmit
with a power $p_k^t=p_k/h_{{ST_k},PR}$.
\item SUs can achieve different (fixed) rates to their own receivers (i.e.,
data rate $r_{SU_k}$ over link $ST_k-SR_k$) with different (fixed)
transmission power (i.e., $p_{SU_k}^t$).
\item SUs have different cost $C_k$ per unit transmission power.
\end{itemize}

%

Note that the parameters $C_k$, $r_{SU_k}$, $p_{SU_k}^t$, and
$h_{ST_k,PR}$ are SU $k$'s private information and are only known to
himself.

We define an $SU$ $k$'s \textbf{payoff} as $\pi_{SU_k}$, which is
the difference between its own transmitted data during time
allocation $t_k$ in Phase III and its cost of power consumption
during Phase II and $t_k$ in Phase III. That is,
\begin{equation}\label{eq:SU_payoff}
\pi_{SU_k}=t_kr_{SU_k}-\left(t_kp^t_{SU_k}+\frac{1}{2}\frac{p_k}{h_{ST_k,PR}}\right)C_k.
\end{equation}
We assume that every SU is willing to use positive transmission time
if it does not need to relay the PU's traffic, i.e.,
$r_{SU_k}-p_{SU_k}^tC_k\geq 0$ for all
$k\in\mathcal{N}$.\footnote{If this is not true for an SU, we can
simply eliminate it from the network.} Notice that
(\ref{eq:SU_payoff}) is increasing in time allocation $t_k$, but is
decreasing in relay power $p_k$.

We can further simplify (\ref{eq:SU_payoff}) by multiplying both
sides by ${2h_{ST_k,PR}}/{C_k}$, which leads to the normalized
payoff
\begin{equation}\label{eq:SU_payoff2}
\tilde{\pi}_{SU_k}:={\pi_{SU_k}}{\frac{2h_{ST_k,PR}}{C_k}}=\frac{2h_{ST_k,PR}(r_{SU_k}-C_kp_{SU_k}^t)}{C_k}t_k-p_k.
\end{equation}
Such normalization does not affect SUs' choice among different relay
powers and time allocations. Thus it will not affect the contract
design introduced later.

To facilitate later discussions, we define an $SU$ $k$'s type as
\begin{equation}\label{eq:type}
\theta_k:=\frac{2h_{ST_k,PR}(r_{SU_k}-C_kp_{SU_k})}{C_k}>0,
\end{equation}
which captures all private information of this SU. A large type
$\theta_k$ means that the SU's own transmission is efficient (a
large $r_{SU_k}$ or a small $p_{SU_k}$), or it has good channel
condition over relay link $ST_k-PR$ (a large channel gain
$h_{ST_k-PR}$), or it has a more efficient battery technology (a
small $C_k$).
\comg{Here, we denote the set of SU types as
$\mathcal{K}=\{1,2,...,K\}$.  Such a definition of $\mathcal{K}$ may confuse
with $\Theta$.  So I move the definition of $\mathcal{K}$ in the below of this column.} With (\ref{eq:type}), we can simplify
$SU$'s normalized payoff in (\ref{eq:SU_payoff2}) as
\begin{equation}\label{eq:SU_payoff3}
\pi_k(p_k,t_k):=\tilde{\pi}_{SU_k}=\theta_kt_k-p_k.
\end{equation}
which is decreasing in PU's received power $p_k$ and increasing in
PU's time allocation $t_k$ to $SU$ $k$.

\com{In the following we seem to use $k$ as SU index. In that case, let us change $n$ into $k$ in the previous paragraphs.}

Since each SU is selfish, a type-$\theta_k$ SU wants to choose relay
power and time allocation to maximize its payoff in
(\ref{eq:SU_payoff3}). Notice that an SU can always choose not to
help the PU and thus receives zero time allocation and zero payoff
(i.e., $t_k=p_k=0$).

\subsection{Contract Formulation under Incomplete
Information}\label{subsec:SU} After introducing PU's utility in
(\ref{eq:PU_utility}) and SUs' (normalized) payoffs in
(\ref{eq:SU_payoff3}), we are ready to introduce the contract
mechanism that resolves the conflicting objectives between the PU
and SUs.

Contract theory studies how economic decision-makers construct
contractual arrangements, generally in the presence of asymmetric (private)
information\cite{ContractTheory}. In our case, the SUs' types are
private information, thus their types are only known to themselves.
The PU does not know the type of each SU, and needs to design a
contract to attract the SUs to participate in cooperative
communications.

To better understand the contract design in this paper, we can
imagine the PU as the employer and SUs as employees in a labor
market. The employer determines the contract, which specifies the
relationship between the employee's performance (i.e., received
relay powers) and reward (i.e., time allocation). If we denote
$\mathcal{P}$ as the set of all possible relay powers and
$\mathcal{T}$ as the set of all possible time allocations, then the
contract specifies a $t\in\mathcal{T}$ for every $p\in\mathcal{P}$.
Each distinct power-time association becomes a contract item. Once a
contract is given, each SU will choose the contract item that
maximizes its payoff in (\ref{eq:SU_payoff3}). The PU wants to
optimize the contract items to maximize its utility in
(\ref{eq:PU_utility}).

We consider $K$ types of SUs with types denoted by the set
$\Theta=\{\theta_1,\theta_2,...,\theta_K\}$. Without loss of
generality, we assume that $\theta_1<\theta_2<...<\theta_K$.
The total number of SUs in type-$\theta_k$ is $N_k$. According to
the revelation principle\cite{GameTheory}, it is enough to consider
the class of contract that enables the SUs to truthfully
\com{report}\rev{reveal} their types. Because of this, it is enough
to design a contract that consists $K$ contract items, one for each
type. The contract can be written as $\Phi=\{(p_k,t_k),\forall
k\in\mathcal{K}\}$ \revh{where $\mathcal{K}=\{1,2,...,K\}$}.

We will derive the optimal contract design for three
information scenarios.
\begin{itemize}
\item \emph{Complete information in Section~\ref{sec:complete}:} This is a benchmark case,
where the PU knows each SU's type. We will compute the maximum
utility the PU can achieve in this case, which serves as an
upper-bound of the PU's achievable utility in the incomplete
information scenarios.

\item \emph{Weak incomplete information in Section~\ref{sec:incomplete1}:} The PU does not know each
SU's type, but has knowledge of the number of each type SUs in the
market (i.e., $N_k$ for type-$\theta_k$ SUs). We will show that the
optimal contract in this case achieves the same maximum PU's utility
as in the complete information benchmark.

\item \emph{Strong incomplete information in
Section~\ref{sec:incomplete2}:} The PU only knows the total number
of SUs ($N$) and the distribution of each type, but does not know the
number of each type ($N_k$). The PU needs to
design a contract to maximize its \emph{expected} utility.
\end{itemize}

Once the PU has determined the contract, the interactions between
the PU and SUs will follow four steps.
\begin{enumerate}
\item The PU broadcasts the contract $\Phi=\{(p_k,t_k),\forall k\in\mathcal{K}\}$ to all
SUs.
\item After receiving the contract, each SU chooses
one contract item that maximizes its payoff and informs the PU
its choice.
\item After receiving all SUs confirmations, the PU informs the involved SUs
(i.e., those choosing positive contract items) the space-time codes
to use in Phase II and the transmission schedule in Phase III. Note
that the length for transmission time for each involved SU is specified by the contract item and  the PU can no longer change.
\item The communications start by following three phases  in Fig.~\ref{fig:network}.
\end{enumerate}

\section{Optimal Contract Design under complete
information: The Benchmark Scenario}\label{sec:complete}

In the complete information scenario, the PU knows the
type of each SU. We
will use the maximum PU's utility achieved in this case as a
baseline to evaluate the performance of the proposed contracts under
incomplete information in Sections~\ref{sec:incomplete1} and
\ref{sec:incomplete2}.  Without
loss of generality, we assume that $N_k\geq 1$ for all type
$k\in\mathcal{K}$.

As the PU knows each SU's type, it can monitor and make sure that
each type of SUs accepts only the contract item designed for that type.
The PU needs to ensure that the SUs have non-negative payoffs
so that they are willing to accept the contract. In other words, the
contract needs to satisfy the following individual rationality
constraint.
\begin{definition}\label{def:IR}
(IR: Individual Rationality): A contract satisfies the
individual rationality constraint if each type-$\theta_{k}$ SU
receives a non-negative payoff by accepting the contract item
for $\theta_{k}$, i.e.,
\begin{equation}\label{eq:IR}
\theta_kt_k-p_k \geq 0, \forall k\in\mathcal{K}.
\end{equation}
\end{definition}


We say a contract is \emph{optimal} if it yields the maximum utility
for the PU under the current information scenario. Different
information scenarios may lead to different optimal contracts.

In the
complete information scenario, an optimal contract maximizes the PU's utility as follows
\comg{Add IR constraints in the optimization function}
\begin{equation}\label{eq:PU_complete}
\begin{aligned}
\max_{\{(p_k,t_k)\geq \boldsymbol{0},\forall k\}} &
\frac{\frac{R^{dir}}{2}+\frac{1}{2}\log\left(1+\frac{\sum_{k\in\mathcal{K}}N_kp_k}{n_0}\right)}{1+\sum_{k\in\mathcal{K}}N_kt_k}, \\
\text{subject to } & \text{ IR Constraints in Eq. } (\ref{eq:IR}).
\end{aligned}
\end{equation}
In this paper, the vector operations are component-wise (e.g, $(p_k,t_k)\geq \boldsymbol{0}$ means that $p_k\geq0$ and
$t_k\geq0$) unless specified otherwise. Then we have the following result.

\begin{lemma}\label{lemma:complete}
In an optimal contract with complete information, each SU receives
zero payoff by accepting the corresponding contract item. In other
words, $t_k\theta_k=p_k$ for any $k\in\mathcal{K}$.
\end{lemma}

\emph{Proof.} We prove by contradiction. Suppose that there exists
an optimal contract item $(p_k,t_k)$ with $\theta_kt_k-p_k>0$. Since
PU's utility in (\ref{eq:PU_complete}) is increasing in $p_k$ and
decreasing in $t_k$, the PU can increase its utility by decreasing
$t_k$ until $\theta_kt_k-p_k=0$. This contradicts with the
assumption that $(p_k,t_k)$ with $\theta_kt_k-p_k>0$ belongs to an
optimal contract, and thus completes the proof.
\hfill$\rule{2mm}{2mm}$

Using Lemma~\ref{lemma:complete}, we can replace $p_k$ by
$\theta_kt_k$ for each $k\in\mathcal{K}$ and simplify the PU's
utility maximization problem in (\ref{eq:PU_complete}) as
\begin{align}\label{eq:PU_optimization_complete}
&\max_{\{t_k\geq 0,\forall
k\}}\frac{\frac{R^{dir}}{2}+\frac{1}{2}\log\left(1+\frac{\sum_{k\in\mathcal{K}}N_k\theta_kt_k}{n_0}\right)}{1+\sum_{k\in\mathcal{K}}N_kt_k}.
\end{align}

%

\begin{theorem}\label{thm:complete}
In an optimal contract with complete information, only the contract
item for the highest type is positive and all other contract items
are zero. That is $(p_K,t_K)>\boldsymbol{0}$, and
$(p_k,t_k)=\boldsymbol{0}$ for any $k<K$.
\end{theorem}

\emph{Proof.} We prove by contradiction. Suppose that there exists
an optimal contract item with $t_k>0$ for the type-$\theta_k$ SUs
and $k<K$. The total time allocation is
$T'=\sum_{k\in\mathcal{K}}N_kt_k$. Then the PU's utility is
\begin{equation}\label{eq:complete_tkk}
u_{PU}^1=\frac{1}{1+T'}\left(\frac{R^{dir}}{2}+\frac{1}{2}\log\left(1+\frac{\sum_{k\in\mathcal{K}}\theta_kN_kt_k}{n_0}\right)\right).
\end{equation}
Next, we show that given a fixed total time allocation $T'$,
allocating positive time only to the highest type SUs (i.e., $N_Kt_K=T'$) achieves a larger utility for the PU as follows
\begin{equation}\label{eq:complete_tK}
u_{PU}^2=\frac{1}{1+T'}\left(\frac{R^{dir}}{2}+\frac{1}{2}\log\left(1+\frac{\theta_KN_Kt_K}{n_0}\right)\right).
\end{equation}
This is because  $\theta_KN_Kt_K=\theta_KT'$ in (\ref{eq:complete_tK}) and
$\sum_{k\in\mathcal{K}}\theta_kN_kt_k<\theta_KT'$ in
(\ref{eq:complete_tkk}), thus (\ref{eq:complete_tK}) is
larger than (\ref{eq:complete_tkk}). This contradicts with the
optimality of the contract, and thus completes the
proof.\hfill$\rule{2mm}{2mm}$

\begin{figure}[tt]
\centering
\includegraphics[width=0.35\textwidth]{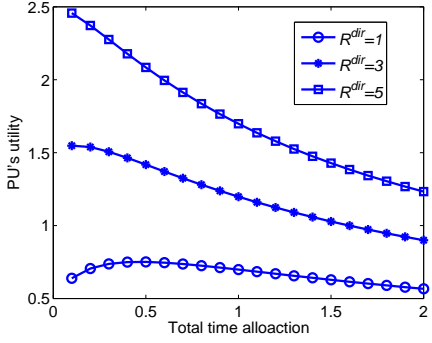}
\caption{PU's utility as a function of the total time allocation
$\tilde{t}_K^*$ for different direct transmission rates $R^{dir}$.}
\label{fig:nonconvexity}
\end{figure}

Intuitively, the highest type SUs can offer the most help to the PU
within the given total time allocation in Phase III.

Using Theorem~\ref{thm:complete}, the optimization problem in
(\ref{eq:PU_optimization_complete}) can be simplified as
\begin{align}\label{eq:complete_optimization}
&\max_{t_K\geq
0}\frac{1}{1+N_Kt_K}\left(\frac{R^{dir}}{2}+\frac{1}{2}\log\left(1+\frac{\theta_KN_Kt_K}{n_0}\right)\right).
\end{align}
Since $N_K$ and $t_K$ always appear as a product in
(\ref{eq:complete_optimization}), we can redefine the optimization
variable as $\tilde{t}_K=N_Kt_K$ and rewrite
(\ref{eq:complete_optimization}) as
\begin{equation}\label{eq:opt_complete}
\max_{\tilde{t}_K>0}\frac{1}{1+\tilde{t}_K}\left(\frac{R^{dir}}{2}+\frac{1}{2}\log\left(1+\frac{\theta_K\tilde{t}_K}{n_0}\right)\right),
\end{equation}
This means the PU's optimal utility does not depend on $N_K$. When
$N_K$ changes, the optimal time allocation per user $t_K^\ast$
changes inversely proportional to $N_K$.

At this point, we have successfully simplified the PU's optimization
problem from involving $2K$ variables $\{(p_k,t_k),\forall
k\in\mathcal{K}\}$ in (\ref{eq:PU_complete}) to a single
variable $\tilde{t}_K$ in (\ref{eq:opt_complete}).

The problem (\ref{eq:opt_complete}) is a non-convex optimization
problem, as shown in Fig.~\ref{fig:nonconvexity} for various
parameters. Although it is difficult to find a closed-form solution,
we can use an efficient one-dimensional exhaustive search algorithm
to find the global optimal solution
$\tilde{t}_K^\ast$\cite{Nonlinear}. We will provide numerical
results in Section~\ref{sec:Numerical}.


\section{Feasible Contracts Under Incomplete Information}\label{sec:feasible}

%

In this
section, we study the necessary and sufficient conditions for a
feasible contract under incomplete information. This will help us
derive optimal contracts in Sections~\ref{sec:incomplete1} and
\ref{sec:incomplete2}.

A feasible contract includes $K$ power-time items such that any
type-$\theta_k$ SU prefers the contract item $(p_k,t_k)$ for its
type to any other contract item. A feasible contract must satisfy
both the individual rationality (IR) constraint in
Definition~\ref{def:IR} introduced in Section~\ref{sec:complete} and
the incentive compatibility (IC) constraint defined next.
\begin{definition}
(IC: Incentive Compatibility): A contract satisfies the
incentive compatibility constraint if each type-$\theta_{k}$ SU
prefers to choose the contract item for $\theta_{k}$, i.e.,
\begin{equation}\label{eq:IC}
\theta_kt_k-p_k\geq \theta_kt_j-p_j, \forall k, j\in \mathcal{K}.
\end{equation}
\end{definition}

In summary, the PU's optimization problem is
\begin{align}\label{eq:PU_optimization}
&\max_{\{(p_k,t_k),\forall k\}}u_{PU}(\{(p_k,t_k),k\in\mathcal{K}\}),\\
&\text{subject\ to}\ \ \theta_kt_k-p_k\geq \theta_kt_j-p_j,
\ \forall k,j\in\mathcal{K},\nonumber\\
& \ \ \ \ \ \ \ \ \ \ \ \ \ \theta_kt_k-p_k\geq 0,\forall k\in\mathcal{K}\nonumber\\
&\ \ \ \ \ \ \ \ \ \ \ \ \ t_k\geq0, p_k\geq 0,\ \forall
k\in\mathcal{K}.\nonumber
\end{align}
The first two constraints correspond to IC and IR, respectively.

\subsection{Sufficient and Necessary Conditions for
Feasibility}\label{subsec:suff_necc_feasible}
%

Next we provide several necessary and sufficient conditions for
the contract feasibility.

\begin{proposition}\label{prop:necc_t_p}
[Necessary condition 1]: For any feasible contract
$\Phi=\{(p_k,t_k),\forall k\}$, we have $p_i>p_j$ if and only if
$t_i>t_j$.
\end{proposition}

\emph{Proof.} We divide the proof into two parts. First, we prove
that if $p_i>p_j$, then $t_i>t_j$. Due to the IC constraint in
(\ref{eq:IC}), we have
$$\theta_it_i-p_i\geq \theta_it_j-p_j,$$
i.e., $$\theta_i(t_i-t_j)\geq p_i-p_j.$$ Since $p_i>p_j$, we
conclude
$$\theta_i(t_i-t_j)\geq p_i-p_j>0,$$
and thus $t_i>t_j$.

Next we prove that if $t_i>t_j$, then $p_i>p_j$. Due to the IC
constraint in (\ref{eq:IC}), we have
$$\theta_jt_j-p_j\geq \theta_jt_i-p_i,$$
which can be transformed to be
$$p_i-p_j\geq \theta_j(t_i-t_j).$$
Since $t_i>t_j$, we conclude
$$p_i-p_j\geq \theta_j(t_i-t_j)>0,$$
and thus $p_i>p_j$. \hfill$\rule{2mm}{2mm}$

Proposition~\ref{prop:necc_t_p} shows that an SU contributing more in
terms of received power at the PU receiver should receive more time
allocation, and vice versa.
From Proposition \ref{prop:necc_t_p}, we have the following
corollary, saying that the same relay powers must have the same time
allocations, and vice versa.

\begin{corollary}
For any feasible contract $\Phi=\{(p_k,t_k),\forall k\}$, we have
$p_i=p_j$ if and only if $t_i=t_j$.
\end{corollary}

Proposition~\ref{prop:necc_theta_t} shows the second necessary
condition for contract feasibility.

\begin{proposition}\label{prop:necc_theta_t}[Necessary Condition 2]:
For any feasible contract $\Phi=\{(p_k,t_k),\forall k\}$, if
$\theta_i>\theta_j$, then $t_i\geq t_j$.
\end{proposition}

\emph{Proof.} We prove by contradiction. Suppose that there exists
$t_i<t_j$ with $\theta_i>\theta_j$. Then we have
\begin{equation}\label{eq:IP_proof}
\theta_it_j+\theta_jt_i>\theta_it_i+\theta_jt_j.
\end{equation}
On the other hand, the feasible contract satisfies the IC
constraints for both type-$\theta_i$ and type-$\theta_j$ SUs, i.e.,
$$\theta_it_i-p_i\geq \theta_it_j-p_j,$$
and
$$\theta_jt_j-p_j\geq \theta_jt_i-p_i.$$
By combining last two inequalities, we have
$$\theta_it_i+\theta_jt_j\geq \theta_it_j+\theta_jt_i,$$
which contradicts with (\ref{eq:IP_proof}). This completes the
proof. \hfill$\rule{2mm}{2mm}$

Proposition~\ref{prop:necc_theta_t} shows that a higher type SU
should be allocated more transmission time. Combined with
Proposition \ref{prop:necc_t_p}, we know that a higher type of SU
should also contribute more in terms of PU's received power.

From Propositions \ref{prop:necc_t_p} and \ref{prop:necc_theta_t},
we conclude that for a feasible contract, all power-time combination
items satisfy
\begin{equation}\label{eq:neccessary}
0\leq p_1\leq p_2\leq ...\leq p_K, \ 0\leq t_1\leq t_2\leq...\leq
t_K,
\end{equation}
with $p_k=p_{k+1}$ if and only if $t_k=t_{k+1}$.

The previous propositions help us obtain Theorem~\ref{prop:suff}
as follows.

\begin{theorem}\label{prop:suff}
[Sufficient and Necessary Conditions]: For a contract
$\Phi=\{(p_k,t_k),\forall k\}$, it is feasible if and only if all
the following three conditions hold:
\begin{itemize}
\item $\mathtt{Contd.a}$: $0\leq p_1\leq p_2\leq ...\leq p_K$ and $0\leq t_1\leq
t_2\leq...\leq t_K$;
\item $\mathtt{Contd.b}$: $\theta_1t_1-p_1\geq 0$;
\item $\mathtt{Contd.c}$: For any $k=2,3,...,K$,
\begin{equation}\label{eq:contdc}
p_{k-1}+\theta_{k-1}(t_k-t_{k-1})\leq p_k\leq
p_{k-1}+\theta_k(t_k-t_{k-1}).
\end{equation}
\end{itemize}
\end{theorem}

We give the proof of Theorem~\ref{prop:suff} in Appendix~\ref{proof:thm1}.
The conditions in Theorem~\ref{prop:suff} are essential to the
optimal contract design under weak and strong incomplete information
in Section~\ref{sec:incomplete1} and Section~\ref{sec:incomplete2}.


\section{Optimal Contract Design under weak incomplete
information}\label{sec:incomplete1}

In this section, we will look at the weak incomplete scenario where
the PU \revh{does not know each SU's type but} only knows the number
of each type (i.e., $N_k$ for any $k\in\mathcal{K}$). Without loss
of generality, we assume that $N_k\geq 1$ for all $k\in\mathcal{K}$.
Different from the complete information case, here the PU cannot
force an SU to accept certain contract item as the PU does not know
the SU's type. Thus we need to consider IC constraint here (while
not in the complete information case in Section~\ref{sec:complete}).


A conceptually straightforward approach to derive the optimal
contract is to solve (\ref{eq:PU_optimization}) directly. Going
through this route, however, is very challenging as
(\ref{eq:PU_optimization}) is a non-convex and involves complicated
constraints.

Here we adopt a sequential optimization approach instead: we first derive
the best relay powers $\{p_k^{\ast}(\{t_k,\forall k\}),\forall k \}$ given
fixed feasible time allocations $\{t_k,\forall k\}$, then derive the
best time allocations $\{t_k^\ast,\forall k\}$ for the optimal
contract, and finally show that there is no gap between the solution
$\{(p_k^{\ast},t_k^\ast),\forall k\}$ obtained from this sequential
approach and the one obtained by directly solving
(\ref{eq:PU_optimization}).

\begin{proposition}\label{prop:incomplete1}
Let $\Phi=\{(p_k,t_k),\forall k\}$ be a feasible contract with fixed
time allocations $\{t_{k},\forall k:0\leq t_1\leq ... \leq t_K\}$.
The optimal unique relay powers satisfy
\begin{align}\label{eq:incomplete_power}
&p_1^{\ast}(\{t_k,\forall k\})=\theta_1t_1,\nonumber\\
&p_k^{\ast}(\{t_k,\forall
k\})=\theta_1t_1+\sum_{i=2}^k\theta_i(t_i-t_{i-1}),\forall
k=2,...,K.
\end{align}
\end{proposition}

The proof of this proposition is given in Appendix~\ref{proof:prop}.
Using Proposition \ref{prop:incomplete1}, we can simplify
%
the PU's optimization problem in (\ref{eq:PU_optimization}) as
\begin{align}\label{eq:incomplete_optimization}
&\max_{\{t_k,\forall k\}}\frac{\frac{R^{dir}}{2}+\frac{1}{2}\log\left(1+\frac{\sum_{k\in\mathcal{K}}N_k(\theta_1t_1+\sum_{i=1}^k\theta_i(t_i-t_{i-1}))}{n_0}\right)}{{1+\sum_{k\in\mathcal{K}}N_kt_k}},\\
&\text{subject\ to},\ \ \  \ \ \ \ \ \ \ \  0\leq t_1\leq ... \leq
t_K.\nonumber
\end{align}
We can further simplify (\ref{eq:incomplete_optimization}) using
Theorem~\ref{thm:incomplete1} below.

\begin{theorem}\label{thm:incomplete1}
In an optimal contract with weak incomplete information, only the
contract item for the highest SU type is positive and all other
contract items are zero. That is $(p_K,t_K)>\boldsymbol{0}$, and
$(p_k,t_k)=\boldsymbol{0}$ for any $k<K$.

\end{theorem}



Using Theorem~\ref{thm:incomplete1}, we can simplify the
optimization problem in (\ref{eq:incomplete_optimization}) further as
\begin{align}\label{eq:incomplete_optimization2}
&\max_{t_K\geq
0}\frac{1}{1+N_Kt_K}\left(\frac{R^{dir}}{2}+\frac{1}{2}\log\left(1+\frac{\theta_KN_Kt_K}{n_0}\right)\right).
\end{align}

Notice that (\ref{eq:incomplete_optimization2}) under weak
incomplete information is the same as
(\ref{eq:complete_optimization}) under complete information.
We thus conclude that our sequential optimization approach (first
over $\{p_k,\forall k\}$ and then over $\{t_k,\forall k\}$) results
in no loss in optimality, as it achieves the same maximum utility as
in the complete information scenario.


To solve problem (\ref{eq:incomplete_optimization2}), we can use an
efficient one-dimensional exhaustive search algorithm to find the
global optimal solution $t_K^\ast$.
We will provide numerical results in Section~\ref{sec:Numerical}.


\section{Optimal Contract Design under strong incomplete
information}\label{sec:incomplete2}

\begin{table*}[tt]
\begin{align}
\max_{\{(p_k,t_k),\forall
k\}}\sum_{n_1=0}^N\sum_{n_2=0}^{N-n_1}...\sum_{n_{K-1}=0}^{N-\sum_{i=1}^{K-2}n_i}\frac{Q_{(n_1,...,n_{K-1},N-\sum_{i=1}^{K-1}n_i)}}{1+\sum_{i=1}^{K-1}n_it_i+(N-\sum_{i=1}^{K-1}n_i)t_K}\left(\frac{R^{dir}}{2}+\frac{1}{2}\log\left(1+\frac{\sum_{i=1}^{K-1}n_ip_i+({N-\sum_{i=1}^{K-1}n_i})p_K}{n_0}\right)\right).\nonumber
(24)
\end{align}
\begin{align}
&\ \ \max_{\{t_k,\forall k\}}\sum_{n_1=0}^N\sum_{n_2=0}^{N-n_1}...\sum_{n_{K-1}=0}^{N-\sum_{i=1}^{K-2}n_i}\frac{{Q_{(n_1,...,n_{K-1},N-\sum_{i=1}^{K-1}n_i)}}\left({\frac{R^{dir}}{2}+\frac{1}{2}\log\left(1+\frac{\sum_{i=1}^{K-1}n_ip_i^\ast(\{t_k,\forall k\})+({N-\sum_{i=1}^{K-1}n_i})p_K^\ast(\{t_k,\forall k\})}{n_0}\right)}\right)}{1+\sum_{i=1}^{K-1}n_it_i+(N-\sum_{i=1}^{K-1}n_i)t_K}.\nonumber \ \ (26)\\
&\text{subject\ to},\ \ \ \ \ \ \ \ \ \ \ \ \ \ \ \ \ \ \ \ \ \ \ \
\ \ \ \ \ \ \ \ \ \ \ \ \ \ \ \ \ \ \ \ \ \ \ \ \ \ \ \ \ \ \ \ \ \
\ \ \ \ \ \ \ \ \ \ 0\leq t_1\leq ...\leq t_K.\nonumber
\end{align}
\hrule
\end{table*}

In this section, we study the strong incomplete information
scenario, where the PU does not know each SU's type and even the
number of each type (i.e., $N_k$ for each $k\in\mathcal{K}$). The
PU only knows the total number of SUs $N$ and the probability $q_k$
\revh{of each SU belonging to type-$\theta_k$ (i.e., $\sum_{k\in\mathcal{K}}q_k=1$).}

Similar to the weak incomplete information scenario, here the PU
needs to consider the IC constraint since it cannot force an SU to
accept a certain contract item. The difference from
Section~\ref{sec:incomplete1} is that here the PU does not know
whether $N_K=0$ or $N_K>0$ for the highest type-$\theta_K$, and thus
the simple approach of only providing a positive contract item for type $\theta_K$
as in Theorem~\ref{thm:incomplete1} may not be optimal. If the PU does that and it turns out
that $N_K=0$ in a particular realization, then there will be no SUs
participating in the cooperative communications.

The right target for the PU is design a contract to maximize the \emph{expected} utility subject to the IC and IR constraints.
As the PU knows the total number of SUs $N$, then the probability
density function of the number of SUs
$\{N_k,\forall k\}$ is
\begin{align}\label{eq:prob_N1NK}
&Q_{(n_1,...,n_{K-1},n_K=N-\sum_{i=1}^{K-1}n_i)}\nonumber\\
&:=\Pr(N_1=n_1,...,N_{K-1}=n_{K-1},N_K=N-\sum_{i=1}^{K-1}N_i)\nonumber\\
&=\frac{N!}{n_1!...n_{K-1}!(N-\sum_{i=1}^{K-1}n_i)!}q_1^{n_1}...q_{K-1}^{n_{K-1}}q_K^{N-\sum_{i=1}^{K-1}n_i}.
\end{align}
\rev{The PU's optimization problem can be written in (24) subject to
the IC ad IR constraints}.

Similar to Section~\ref{sec:incomplete1}, here we adopt a sequential
optimization approach: we first derive the optimal relay powers
$\{p_k^{\ast}(\{t_k,\forall k\}),\forall k\}$ with fixed feasible
time allocations $\{t_k,\forall k\}$, then derive the optimal time
allocations $\{t_k^\ast,\forall k\}$ for the optimal contract. The
difference is that optimality is no longer guaranteed here as
explained later.

\begin{proposition}\label{prop:incomplete2}
Let $\Phi=\{(p_k,t_k),\forall k\}$ be a feasible contract with fixed
time allocations $\{t_{k},\forall k:0\leq t_1\leq ... \leq t_K\}$,
then the unique optimal relay powers satisfy
\begin{align}
&p_1^{\ast}(\{t_k,\forall k\})=\theta_1t_1,\nonumber\\
&p_k^{\ast}(\{t_k,\forall
k\})=\theta_1t_1+\sum_{i=2}^k\theta_i(t_i-t_{i-1}),\forall
k=2,...,K.\nonumber \ \ (25)
\end{align}
\end{proposition}

Notice that Proposition~\ref{prop:incomplete2} under strong
incomplete information is actually the same as
Proposition~\ref{prop:incomplete1} under weak incomplete
information. Proposition~\ref{prop:incomplete2} can be proved in a
similar manner as Proposition~\ref{prop:incomplete1}, using the fact
that PU's expected utility is increasing in $\{p_k,\forall k\}$. The
proof is given in Appendix~\ref{proof:propincomplete2}. Based on
Proposition~\ref{prop:incomplete2}, we can simplify the PU's
optimization problem in (24) as (26).

\com{(23) and (24) are used after (25) and (26), which are confusing. Please check the manual for IEEEtran style file, which explains how to generate the right equation number sequencing for two-column equations.}

Note that (26) is a non-convex optimization problem and all $K$
variables are coupled in the objective function. Furthermore, the
number of terms in the objective increases exponentially with the
number of types $K$. Thus it is hard to solve efficiently.

Next, we propose a low computation complexity approximate algorithm,
Decompose-and-Compare algorithm, to compute a close-to-optimal
solution to (26) efficiently. In this algorithm, we will compare $K$
simple candidate contracts, and pick the one that yields the largest
utility for the PU. There are two key ideas behind this heuristic
algorithm.
\begin{itemize}
\item \emph{One positive contract item per contract:} \rev{In each of the $K$ individually optimized
candidate contracts, there is only one positive contract item (for
one or more types). For the contract item in the $k$th candidate
contract, this positive contract item is offered to SUs with types
equal to or larger than type-$\theta_k$. In other words,
optimization of each candidate contract only involves a scalar
optimization, instead of $K$ variables as in (26).}

\item \emph{A balance between efficiency and uncertainty:}
\rev{Among $K$ optimized candidate contracts, we will pick the best one that achieves the best trade-off between efficiency and uncertainty (i.e., maximizing the PU's expected utility).
Under strong
incomplete information, it is not clear which type is the highest among all SUs existing in a particular network realization.
If a candidate contract offers the same positive contract item for
types equal to or larger than $\theta_{k}$, then all SUs in these
types will choose to accept that contract item. The corresponding
SU's payoff is increasing in type, i.e., a type $\theta_{k}$ SU
receives zero payoff and a type-$\theta_{K}$ SU receives the maximum
payoff. Thus choosing a candidate contract with a threshold
$\theta_{k}$ too low will give too much payoffs to the SUs (and thus
reduce the PU's expected utility), but choosing a candidate contract
with a threshold too high might lead to the undesirable case that no
SUs will be able to participate. This requires us to examine all
possibilities (i.e., $K$ candidate contracts) and pick the one with
the best performance.}
%
\end{itemize}

The Decompose-and-Compare algorithm is as follows.
\begin{enumerate}
\item \emph{Decomposition:} We construct the $K$ candidate
contracts as follows. For the $k$th candidate contract,
\begin{itemize}
\item Only offers the same contract item $t_{k}> 0$ to
SUs with a  type equal to or larger than type $\theta_{k}$, and zero for the rest of the tyes.
\item Computes the optimal $t_{k}^\ast$ that maximizes PU's
expected utility in (26).
\end{itemize}
\item \emph{Comparison:} After determining the above $K$ contract candidates, we choose the one that gives PU the highest expected utility.
\end{enumerate}

In Section~\ref{sec:Numerical}, we will use numerical results to
show that the proposed Decompose-and-Compare algorithm achieves a
performance very close to the optimal solution to (26) in most
cases.


%


\section{Numerical Results}\label{sec:Numerical}

Here we use numerical results to show how the PU designs the optimal
contract in different information scenarios.

\subsection{Complete and Weak Incomplete Information Scenarios}\label{subsec:numerical1}

As shown in Section~\ref{sec:complete} and
Section~\ref{sec:incomplete1}, the optimal contract is the same for
complete and weak incomplete information scenarios. By examining the
PU's optimization in (\ref{eq:incomplete_optimization2}) that
applies to both scenarios, we have the following observations.

\begin{observation}\label{ob:PU_complete}
The PU's optimal utility increases in the highest SU type-$\theta_K$ and the PU's direct transmission rate $R^{dir}$.
\end{observation}


%

\begin{figure}[tt]
\centering
\includegraphics[width=0.35\textwidth]{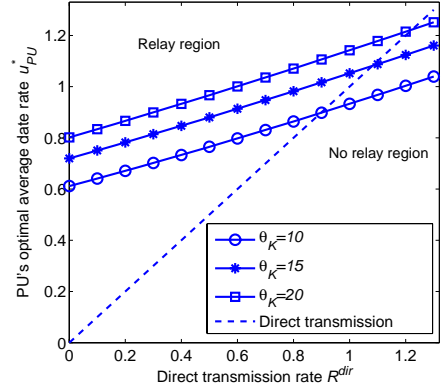}
\caption{PU's optimal utility $u_{PU}^\ast$ as a function of the
PU's direct transmission rate $R^{dir}$ and the highest type
$\theta_K$. The dotted baseline with $45\textordmasculine$ divides
the figure into two regions: Relay Region and No Relay Region.}
\label{fig:PU_DT}
\end{figure}

Figure~\ref{fig:PU_DT} shows PU's utility achieved under the optimal
contract, which is increasing in both $\theta_K$ and direct
transmission rate $R^{dir}$.\footnote{Without loss of generality, we
can normalize $n_0$ to be $1$ in the rest of this paper. Then
$\theta_k$ can be viewed as the ratio $\theta_k/n_0$.} The dotted
baseline denotes the rate $R^{dir}$ achieved by direct transmission
only. As $R^{dir}$ increases, the PU has less incentive to share
spectrum with the SUs. When $R^{dir}$ is very large, the PU chooses
not to use SUs at all, which corresponds to \emph{No Relay Region}
in Fig.~\ref{fig:PU_DT} (where the three curves with different
$\theta_K$'s are below the baseline).


\rev{For the rest of the numerical results, we will only examine the PU's choice of optimal contract, without reiterating the need to compare with $R^{dir}$ and choose direct transmission only if needed.}

\begin{figure}[tt]
\centering
\includegraphics[width=0.4\textwidth]{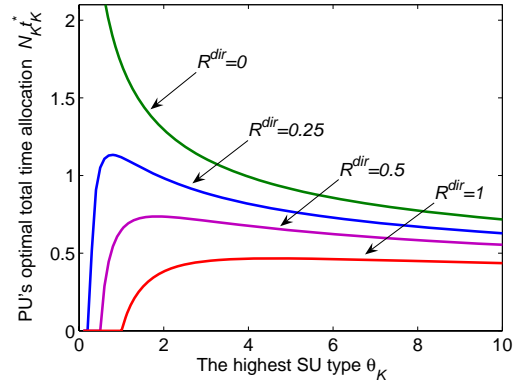}
\caption{PU's optimal total time allocation $N_Kt_K^\ast$ as a
function of the PU's direct transmission rate $R^{dir}$ and the
highest type $\theta_K$.} \label{fig:PU_time_DT}
\end{figure}

\begin{observation}\label{ob:PU_time_DT}
Figure \ref{fig:PU_time_DT} shows that  the PU's optimal total time
allocation to the highest type-$\theta_K$ SUs, $N_Kt_K^\ast$,
decreases in PU's direct transmission rate $R^{dir}$. When direct
transmission rate is zero (i.e., $R^{dir}=0$), the PU's optimal
total time allocation is strictly decreasing in $\theta_K$; when
direct transmission rate is positive (i.e., $R^{dir}>0$), the total time allocation first increases in
$\theta_K$ and then decreases in $\theta_K$.
\end{observation}

When direct transmission rate $R^{dir}=0$, the PU can
only rely on SUs for transmissions and will always allocate positive transmission time
to the highest type-$\theta_K$ SUs. \rev{If we look at the PU's utility
in (\ref{eq:incomplete_optimization2}) with $R^{dir}=0$,
the logarithmic term 
$\log(1+\theta N_Kt_K^\ast)$ plays a more important role than PU's
transmission time ratio $\frac{1}{1+N_Kt_K^\ast}$ in this case. When
$\theta_K$ is small, the PU needs to allocate a large amount time to the SUs to achieve its desirable rate. When $\theta_K$ becomes large, the PU can reach a high
relay rate by allocating less transmission time to the SUs.
This explains why we observe a decrease of $N_Kt_K^\ast$ in $\theta_K$.
Appendix~\ref{proof:ob} provides a rigorous proof of
Observation~\ref{ob:PU_time_DT} under $R^{dir}=0$.}

When direct transmission rate $R^{dir}>0$ (the lower three curves in Fig.~\ref{fig:PU_time_DT}),
the PU has less incentive to allocate transmission time to the SUs
especially when the highest SU type-$\theta_K$ is small. As
$\theta_K$ becomes large, PU is willing to allocate more time in
exchange of efficient help from SUs. As $\theta_K$ becomes very
large, the PU only needs to allocate a small amount of time to the
SUs in order to obtain enough relay help. The above analysis
together explain why the lower three curves in Fig.~\ref{fig:PU_time_DT}
first increase and then decrease in $\theta_K$.

%

\begin{figure*}
 \hspace{-0.5cm}
  \begin{minipage}[t]{0.33\linewidth}
    \centering
\includegraphics[width=1\textwidth]{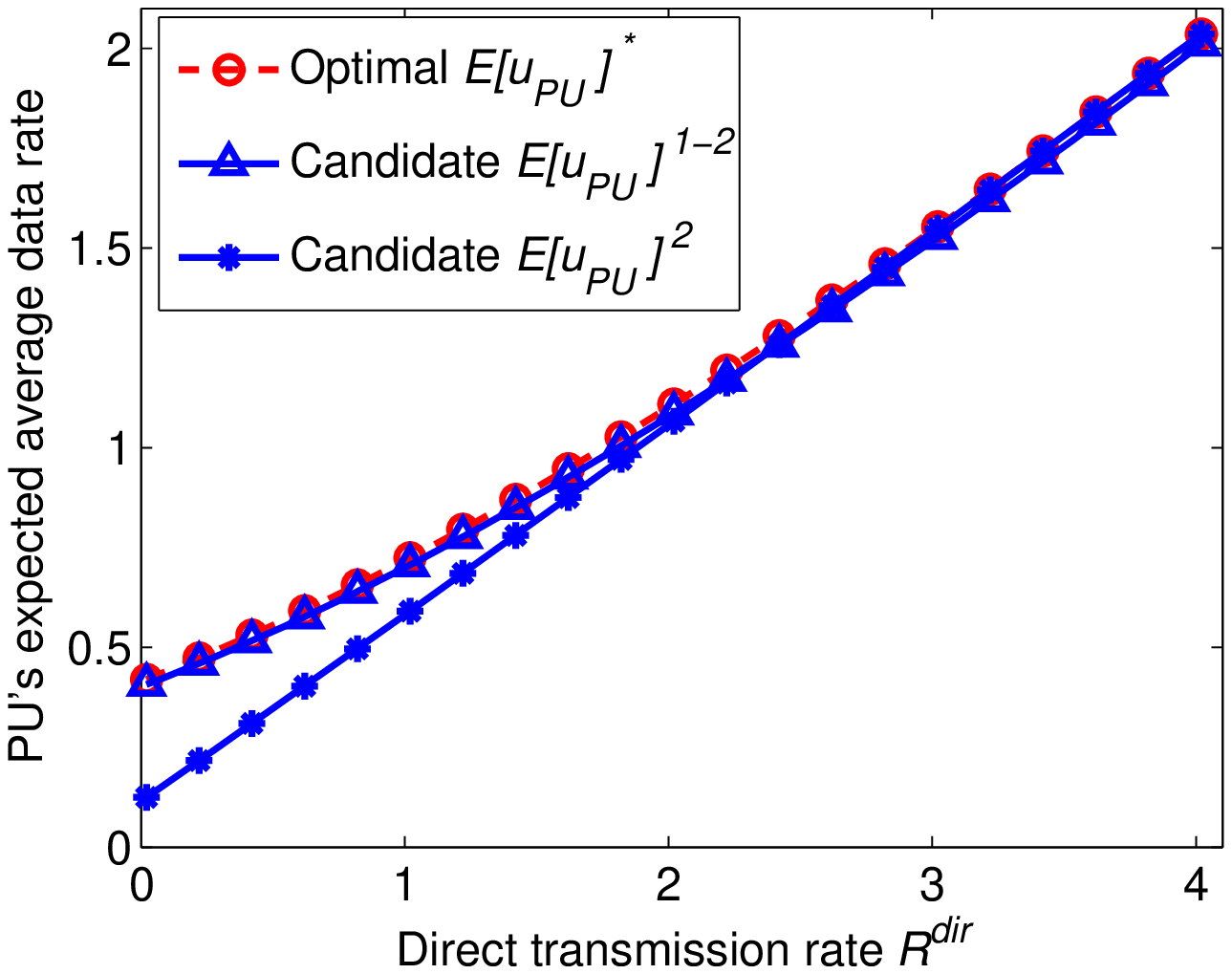}
\caption{Comparison among PU's optimal expected utility using
optimal exhaustive search method ($E[u_{PU}]^\ast$) and the two
candidate contracts of the Decompose-and-Compare algorithm as a
function of the PU's direct transmission rate $R^{dir}$. Other
parameters are $q_1=0.9$, $N=2$, $\theta_1=4$, and $\theta_2=10$.}
\label{fig:approximated1}
    \end{minipage}
  \begin{minipage}[t]{0.01\linewidth}~
  \end{minipage}
  \begin{minipage}[t]{0.33\linewidth}
    \centering
\includegraphics[width=1\textwidth]{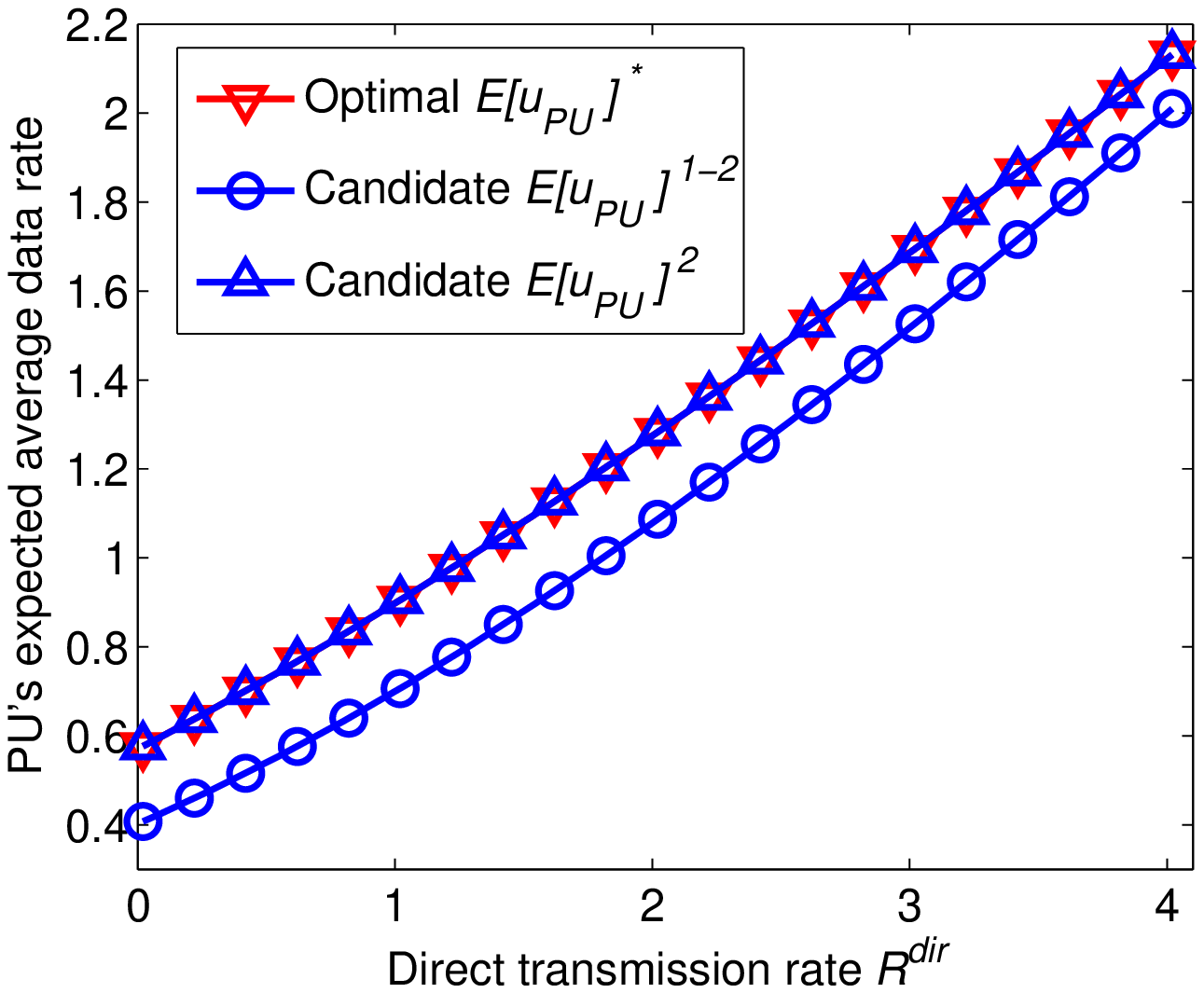}
\caption{Comparison among PU's optimal expected utility using
optimal exhaustive search method ($E[u_{PU}]^\ast$) and the two
candidate contracts of the Decompose-and-Compare algorithm as a
function of the PU's direct transmission rate $R^{dir}$. Other
parameters are $q_1=0.5$, $N=5$, $\theta_1=4$, and $\theta_2=10$.}
\label{fig:approximated3}
  \end{minipage}
  \begin{minipage}[t]{0.01\linewidth}~
  \end{minipage}
  \begin{minipage}[t]{0.33\linewidth}
    \centering
\includegraphics[width=1\textwidth]{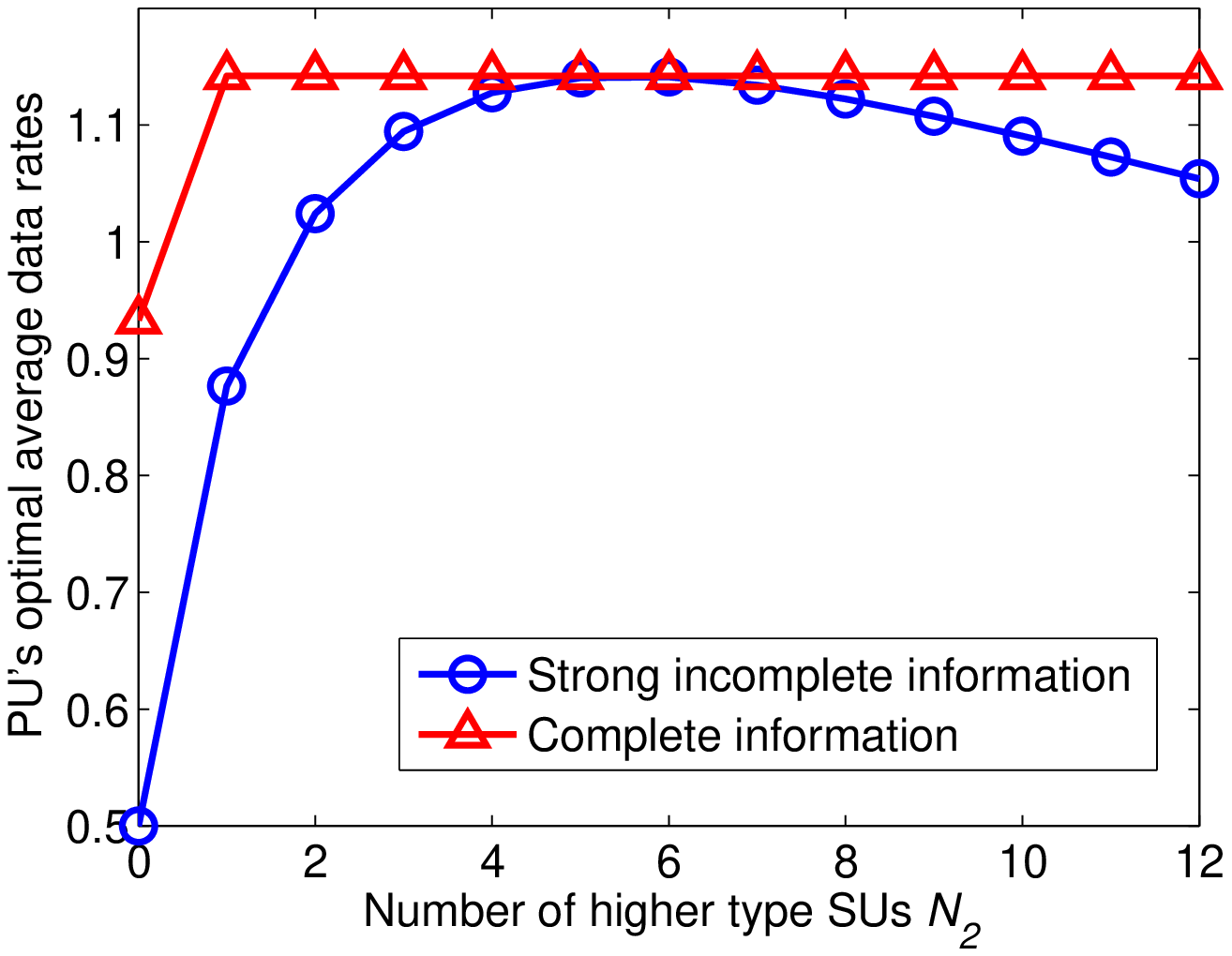}
\caption{Comparison among PU's optimal utilities for any SU number
realization under different information scenarios. Other parameters
are $q_1=0.5$, $N=12$, $R^{dir}=1$, $\theta_1=10$, and
$\theta_2=20$.} \label{fig:comparison_information}
\end{minipage}%
\end{figure*}


\subsection{Strong Incomplete Information Scenario}\label{subsec:numerical2}
Here we show how PU chooses the optimal contract to maximize its
expected utility. As a performance benchmark, we first compute the
optimal solution to the PU's expected utility maximization problem
in (26) via a $K$-dimension exhaustive search. We denote the
corresponding optimal solution as $E[u_{PU}]^\ast$. Notice that
$E[u_{PU}]^\ast$ is often smaller than the maximum utility achieved
under complete information. The performance gap is due to the strong
incomplete information. Next, we will compare the PU's expected
utility achieved by  the proposed Decompose-and-Compare algorithm
with $E[u_{PU}]^\ast$.

For illustration purposes, we consider only two types of SUs:
$\theta_1<\theta_2$. The PU only knows the total number of SUs $N$
and the probabilities $q_1$ and $q_2$ of two types
with $q_1+q_2=1$.

In the Decompose-and-Compare algorithm, we first
consider two candidate contracts. The first candidate contract
optimizes the same positive contract item $t_1=t_2>0$ for both
types. The PU's corresponding maximum expected utility is
$E[u_{PU}]^{1-2}$. The second candidate contract optimizes the
positive contract item $t_2>0$ and chooses $t_1=0$. The PU's
corresponding maximum expected utility is $E[u_{PU}]^2$. Then we
pick the candidate contract that leads to a larger PU's expected utility as the solution of the Decompose-and-Compare algorithm.

To evaluate the performance of this approximate algorithm, we
consider two different parameter regimes.

\subsubsection{Large $q_1^N$}
This means that the probability that all SUs belong to the low
type-$\theta_1$ is large. This happens when the total number of SUs
$N$ is small and probability $q_1$ is large.
Figure~\ref{fig:approximated1} shows the PU's expected utility
obtained from the Decompose-and-Compare algorithm
($\max(E[u_{PU}]^{1-2},E[u_{PU}]^{2})$) and the optimal exhaustive
search method ($E[u_{PU}]^\ast$) as functions of PU's direct
transmission rate $R^{dir}$. We can see that the candidate contract that  offers the same positive
contract items to both types (i.e., $E[u_{PU}]^{1-2}$) achieves a
close-to-optimal performance with all values of $R^{dir}$ simulated here.
This is because very often the PU needs to rely on the low type-$\theta_{1}$ SUs
to relay its traffic.

We also notice that the candidate contract that offers a positive
contract item to the high type SU (i.e., $E[u_{PU}]^{2}$) also
achieves a close-to-optimal performance (even larger than
$E[u_{PU}]^{1-2}$) when $R^{dir}$ is large. This is because the PU
with a large $R^{dir}$ relies less on the SUs, and will have more incentive to employ only the
high type SUs when they are available.

\subsubsection{Small $q_1^N$} This means that the probability
that at least one SU belongs to the high type-$\theta_2$ (i.e.,
$1-q_1^N$) is large. Figure~\ref{fig:approximated3} shows the PU's
expected utility obtained from Decompose-and-Compare algorithm and
the optimal exhaustive search method as a function of $R^{dir}$. We
can see that $E[u_{PU}]^{2}$ is always better than
$E[u_{PU}]^{1-2}$ and achieves a close-to-optimal performance under
all choices of $R^{dir}$. This is because very often the PU can rely
on the high type SUs only to relay its traffic.

\begin{observation}
The performance of the proposed Decompose-and-Compare algorithm
achieves a close-to-optimal performance (i.e., less than $2\%$
according to Fig.~\ref{fig:approximated1} and
Fig.~\ref{fig:approximated3}) under the strong incomplete
information.\footnote{Due to the page limit, we only discuss
two-type case here. We have similar numerical results for cases of
more than two types.}
\end{observation}

\com{Where do we observe the $5\%$ difference?}

\rev{Next we study how the strong incomplete information reduces PU's utility comparing with the complete information benchmark.
First, we note that PU's contract in the strong incomplete information scenario does not depend on $N_{1}$ and $N_{2}$, as the PU targets at optimizing the expected rate and does not know the realization information.} However, the actual PU's utility (not the expected value) does depend on $N_{1}$ and $N_{2}$.
Figure~\ref{fig:comparison_information} shows the PU's utility under
different information scenarios and different user number
realizations (i.e., any realization of $N_1$ and $N_2$ for a fixed
$N_1+N_2=N$). Here, $q_1^N$ is small and the curve with strong
incomplete information corresponds to offering $t_{2}>0$ and
$t_{1}=0$ (as in Fig.~\ref{fig:approximated3}).\rev{ The (very close
to) optimal contract under strong incomplete information can be
obtained by using the Decompose-and-Compare algorithm.} The optimal
contract under complete information changes as $N_1$ and $N_2$
change.

\begin{observation}
Figure~\ref{fig:comparison_information} shows that the PU's optimal
utility under strong incomplete information achieves the maximum
value (close to the one under complete information) when the
realized SU numbers is close to the expected value  (i.e., $N_1=N_2=6$ in this example as $q_{1}=q_{2}=0.5$).

\end{observation}

\rev{In Fig.~\ref{fig:comparison_information}, the largest performance gap between the two curves
happens when $N_2=0$. In this case, the optimal contract under complete information satisfies $t_{1}>0$ and $t_{2}=0$, as there are no type-$\theta_{2}$ users.
However, the PU under strong incomplete information chooses $t_{1}=0$ and $t_{2}>0$ to maximize the PU's expected utility. Such mismatch means that the PU under strong
incomplete information has no SUs serving as relays, and can only
achieve a utility equal to half of the direct transmission rate (only in
Phase~I). However, this parameter setting only happens with a very small probability
$q_1^N=0.5^{12}\approx2.4\times10^{{-4}}$.}

\rev{A more meaning comparison is the PU's \emph{average} utility
loss due to strong incomplete information. We can first compute the
PU's average utility under strong incomplete information, which is
the optimal objective of the PU's expected utility maximization
problem in (26) (via an $K$-dimension exhaustive search). Then we
can compute the PU's average utility under complete information by
calculating the weighted sum (weighted by the probability of each
parameter $(N_{1},N_{2})$) of the 13 values on the upper curve in
Fig.~\ref{fig:comparison_information}. In this example, the ratio is
$0.9874$. This means that the PU's average utility loss due to
strong incomplete information is very small (i.e., less than
$1.3\%$).}

%


\section{Related Work}
\label{sec:RelatedWork}

\subsection{Cognitive Radio and Dynamic Spectrum Sharing}

There are several comprehensive surveys on cognitive radio and
dynamic spectrum sharing \cite{Akyildiz-survey, Marshall1, Mahanen, buddhikot-sur}. Here we will only highlight some recently proposed models. In \cite{HuangLiuDing2008}, Huang \emph{et al.} proposed a distributed spectrum access scheme that enables multiple SUs to collectively protect the PU while adapting to changes in PU activity patterns.  In \cite{cr4}, Xing \emph{et
al.} considered dynamic spectrum sharing with quality of service
guarantee for each secondary link under interference
temperature constraint. In \cite{spshar1}, Cao \emph{et al.}
introduced a distributed spectrum management architecture, where
nodes share spectrum resource fairly by making independent actions
following spectrum rules. In \cite{Mahonen2}, Mehonen \emph{et al.}
discussed how to precisely
model spectrum maps with spatial statistics and random fields, which
is crucial for dynamic spectrum sharing. In \cite{Mahonen3},
Mehonen and Petrova \emph{et al.} further proposed a topology engine
to perform the collecting and processing of spatial information, and
discussed both technical and architectural issues in enabling such
an approach. In \cite{buddhikot}, Buddhikot discussed the potential
benefits of applying cognitive radio technologies in cellular
networks. The above results focused on the technical aspect
of dynamic spectrum sharing without considering the
issue of economic incentives.

\subsection{Spectrum Trading for Dynamic Spectrum Sharing}

Recent years have witnessed a growing body of literatures on the
economic aspect of dynamic spectrum sharing. In \cite{Pierre},
Pierre \emph{et al.} studied the problem of how to determine the
right balance between exclusive use (typically market mechanisms
approach) and licence-exempt use of spectrum. In \cite{nancy}, Lehr
and Jesuale examined the economic, policy, and market challenges of
enabling spectrum pooling, which is considered as the first step
toward dynamic spectrum accessing/sharing. In \cite{lehr}, Chapin
\emph{et al.} investigated time-limited leases for innovative radios
such as cognitive radio and dynamic spectrum access from both
technical and policy perspectives. In \cite{Ballon}, Delaere and
{Ballon} discussed the multi-level standardization and business
models for cognitive radio. In \cite{cr5}, Hwang \emph{et al.}
discussed the feasibility of cognitive radio based network and
application scenarios from the aspects of regulation, policy and
market structure. In \cite{peha}, Peha discussed policies that can
enable or facilitate the use of many spectrum-sharing arrangements.


Market-driven {spectrum
trading} (e.g., \cite{spectrum-trading2}\cite{Weiss})
is a
promising paradigm to address the incentive issue in dynamic
spectrum sharing. 
We can classify spectrum trading models into two types:
money-exchange and resource-exchange. In the former type, SUs pay
PUs in the form of (virtual) money for the usage of spectrum (e.g.,
\cite{pricing1,pricing4,lixy,pricing5,JianweiHuang,
SorabhGandhi,gao2,eBay,TRUST,gao1,contract2}); in the latter type,
SUs provide communication resources (e.g., the transmission power in
our model) for PUs' transmissions in exchange for the usage of
spectrum (e.g., \cite{cocr1,cocr2,cocr3,cocr4}).

There has been extensive research on the money-exchange spectrum
trading model, often in the form of pricing, auction, and contract.
\emph{Pricing} is often used when the seller knows precisely the
value of the resource being sold. In \cite{pricing1}, Kloeck
\emph{et al.} proposed an integrated pricing, allocating, and
billing system for cognitive radio networks. In \cite{pricing4},
Wang \emph{et al.} proposed a joint power and spectrum allocation scheme
using a distributed pricing strategy to improve the network's
performance. In \cite{pricing5}, Niyato \emph{et al.} proposed three
different pricing models for cognitive radio networks with different
objectives.

\emph{Contract} is more effective in the case where the seller only
knows limited information (e.g., distribution) of the buyers'
valuation of the resource. By motivating the buyers truthfully
reveal their private valuations, the seller can
allocate the resource to maximize its own benefit or the social
efficiency. In \cite{gao1}, Gao \emph{et al.} proposed a
quality-price contract for the spectrum trading in a monopoly
spectrum market. In \cite{contract2}, Kalathil \emph{et al.}
proposed a contract-based spectrum sharing mechanism to avoid
possible manipulating in auction.

When the seller has no knowledge about the value of the resource
being sold, \emph{Auction} becomes an effective approach. By letting
the bidders bid for the resource in a truthful manner, the seller
can efficiently allocate the resource without knowing the value
before hand. In \cite{JianweiHuang}, Huang \emph{et al.} proposed
two divisible auction mechanisms for power allocation in spectrum
sharing to achieve efficiency and fairness, respectively. In
\cite{lixy}, Li \emph{et al.} proposed several truthful
(strategy-proof) spectrum auction mechanisms to achieve the
efficiency closed to social optimal. In \cite{SorabhGandhi}, Gandhi
\emph{et al.} proposed a real-time spectrum auction framework to
distribute spectrum among users under interference constraints. In
\cite{gao2}, Gao \emph{et al.} proposed a multi-shot spectrum
auction mechanism to achieve social optimal
efficiency in dynamic spectrum sharing. In \cite{eBay} and
\cite{TRUST}, Zheng \emph{et al.} proposed truthful single-side
spectrum auction and truthful double spectrum auction,
respectively, both considering spectrum reuse among users. Wang \emph{et al.} in \cite{WangXuXu2010} proposed a general framework for truthful online double auction for spectrum sharing.

\begin{table}[t]
\renewcommand{\arraystretch}{1.5}
\centering
\caption{A summary of spectrum trading literatures}
\label{summary-trading}
\begin{tabular}{|c||c|c|}
\hline
\begin{tabular}{c}
{
\renewcommand{\arraystretch}{1.1}
\hspace{-5mm}
\begin{tabular}{c}
                        \textbf{Network } \\
                        \textbf{Information}                     \end{tabular}
\hspace{-5mm}
}
\\
\end{tabular} &
\textbf{Money-Exchange}  &  \textbf{Resource-Exchange} \\
\hline 
 \textbf{Complete} & Pricing: \cite{pricing1,pricing4,pricing5}  & Stackelberg:  \cite{cocr1,cocr2,cocr3,cocr4} \\
\hline 
 \textbf{Incomplete} & {
\renewcommand{\arraystretch}{1.1}
\begin{tabular}{c}
                        Contract: \cite{gao1,contract2} \\
                        Auction: \cite{JianweiHuang,lixy,SorabhGandhi,gao2,eBay,TRUST,WangXuXu2010}                       \end{tabular}}
    & Contract: This paper \\
\hline
\end{tabular}
\end{table}

Money-exchange spectrum trading is most effective when PUs have some temporarily unused spectrum. However, when PUs' own demands are high or the primary channels' capacities are low (e.g., due to shadowing and deep fading), there will be hardly any resource left for sale. In this case, resource-exchange spectrum trading can be a better choice. \emph{Cooperative spectrum sharing} is an effective form of resource-exchange spectrum trading \cite{cocr1,cocr2,cocr3,cocr4}, wherein PUs utilize SUs as cooperative relays. Such cooperation can significantly improve PUs' data rate and thus can free up spectrum resources for SUs. Existing cooperative spectrum sharing mechanisms are usually based on Stackelberg game formulations with complete information \cite{cocr1,cocr2,cocr3,cocr4}. In this paper, we consider the cooperative spectrum sharing under \emph{incomplete information}, and propose a contract-based cooperative spectrum sharing mechanism. As far as we know, this is the first work considering the cooperative spectrum sharing under incomplete information.

We summarize the key literatures of spectrum trading in Table \ref{summary-trading}.


\section{Conclusion}
\label{sec:Conclusion}

We study the cooperative spectrum sharing between one PU and multiple SUs, where the SUs' types are private information. We model the network as a monopoly market, in which the PU offers a contract  and the each SU selects the best contract item according to his type.
We study the optimal contract designing for multiple information scenarios.
We first provide the necessary and sufficient conditions for feasible contracts under any incomplete information. 
For the weak incomplete information scenario, we derive the optimal contract that achieves the same PU's utility as in the complete information benchmark.
For the strong incomplete information scenario, we propose a Decompose-and-Compare algorithm that achieves a close-to-optimal PU's expected utility. 
Both the PU's average utility loss due to the suboptimal algorithm and the strong incomplete information are small in our numerical example (less than $5\%$ and $1.3\%$ in our numerical results with two SU types).

\rev{This work represents a small step towards establishing a general framework of understanding incomplete information in dynamic spectrum sharing. As the next step plan, we will consider more incomplete information structures: (1) the PU does not know the distribution of SUs' types, and (2) each SU knows other SUs' types but PU does not know (a similar setting with a different application has been studied in \cite{addition}). We also want to understand how to design contracts in a market with multiple PUs and multiple SUs.} 


\appendix
\subsection{Proof of Theorem \ref{prop:suff}}
\label{proof:thm1}
\emph{Proof.} We use mathematical induction to prove this
proposition. Let us denote $\Phi(n)$ as a subset of $\Phi$ which
contains the first $n\leq K$ power-time combinations in $\Phi$
(i.e., $\Phi(n)=\{(p_k,t_k)|k=1,...,n\}$).

We first show that $\Phi(1)$ is feasible. Since there is only one SU
type, the contract is feasible if it satisfies IR constraint in
(\ref{eq:IR}). Due to $\mathtt{Contd.b}$ in
Proposition~\ref{prop:suff}, IR constraint is satisfied and
$\Phi(1)$ is feasible.

We then show that if contract $\Phi(k)$ is feasible, then
$\Phi(k+1)$ with introduction of new item $(p_{k+1},t_{k+1})$ is
also feasible. To achieve this, we only need to verify that (I) for
the new type-$\theta_{k+1}$ SU(s), the IC and IR constraints for
this type are satisfied, i.e.,
\begin{equation}\label{eq:add_k+1}
\begin{cases} \; \theta_{k+1}t_{k+1}-p_{k+1}\geq \theta_{k+1}t_i-p_i,\ \forall
i=1,...,k\\
\; \theta_{k+1}t_{k+1}-p_{k+1}\geq 0,
\end{cases}
\end{equation}
and (II) for original types $\theta_1,...,\theta_k$, the IC
constraints are still satisfied in the presence of new type
$\theta_{k+1}$,i.e.,
\begin{equation}\label{eq:add_k+1_original}
\theta_it_i-p_i\geq \theta_it_{k+1}-p_{k+1}, \forall i=1,...,k.
\end{equation}
Note that there is no need to check the IR constraints for original
types after introducing a new type, because original SUs'
non-negative payoffs sticking to their original type items will not
change.
\subsubsection{Proof of (I) in (\ref{eq:add_k+1})}
First, we prove the IC constraint for type $\theta_{k+1}$. Since
contract $\Phi(k)$ is feasible, the IC constraint for a
type-$\theta_i$ SU must hold, i.e., $$\theta_kt_i-p_i\leq
\theta_kt_k-p_k, \forall i=1,...,k.$$ Also, the right inequality of
(\ref{eq:contdc}) can be transformed to
$$p_{k+1}\leq p_k+\theta_{k+1}(t_{k+1}-t_k).$$ By combining the
above two inequalities, we conclude
\begin{equation}\label{eq:proof_I}
\theta_kt_i-p_i+p_{k+1}\leq
\theta_kt_k+\theta_{k+1}(t_{k+1}-t_k),\forall i=1,...,k.
\end{equation}
Notice that $\theta_{k+1}>\theta_k$ and $t_k\geq t_i$ in
$\mathtt{Contd.a}$. By using ID property, we have
$$\theta_{k+1}t_k-\theta_{k+1}t_i\geq \theta_kt_k-\theta_kt_i.$$
By substituting this inequality into (\ref{eq:proof_I}), we conclude
$$\theta_{k+1}t_{k+1}-p_{k+1}\geq \theta_{k+1}t_i-p_i,$$
which is actually the IC constraint for type $\theta_{k+1}$.

Next, we show the IR constraint for type $\theta_{k+1}$. Due to IC
constraint for type $\theta_{k+1}$ which we just proved, we have
$$\theta_{k+1}t_{k+1}-p_{k+1}\geq \theta_{k+1}t_i-p_i,\forall i=1,...,k.$$
Since $\theta_{k+1}>\theta_i$, we can further derive
$$\theta_{k+1}t_{k+1}-p_{k+1}\geq \theta_it_i-p_i,$$
where the right-hand side expression is non-negative due to IR
constraint of type $\theta_i$ in the feasible contract $\Phi(k)$.

\subsubsection{Proof of (II) in (\ref{eq:add_k+1_original})}
Since contract $\Phi(k)$ is feasible, the IC constraint for type
$\theta_i$ holds, i.e.,
$$\theta_it_k-p_k\leq \theta_it_i-p_i, \forall i=1,...,k.$$
Also, we can transform the left inequality of (\ref{eq:contdc}) to
$$p_k+\theta_k(t_{k+1}-t_k)\leq p_{k+1}.$$
By combining the above two inequalities, we conclude
\begin{equation}\label{eq:proof_II}
\theta_it_k+\theta_k(t_{k+1}-t_k)\leq \theta_it_i-p_i+p_{k+1}.
\end{equation}
Note that $\theta_k\geq \theta_i$ and $t_{k+1}\geq t_k$ in
$\mathtt{Contd.a}$. By using ID property, we have
$$\theta_kt_{k+1}-\theta_kt_k\geq \theta_it_{k+1}-\theta_it_k.$$
By combining the above two inequalities, we conclude
$$\theta_it_{k+1}-p_{k+1}\leq \theta_it_{i}-p_i,$$
which is actually the IC constraint for type $\theta_i$.

It is easy to check that the sufficient conditions in
Proposition~\ref{prop:suff} are also the necessary conditions for a
feasible contract. Specifically, $\mathtt{Contd.a}$ is the same as
necessary conditions summarized in (\ref{eq:neccessary}).
$\mathtt{Contd.b}$ is same as the necessary IR constraint for the
lowest type $\theta_1$ in a feasible contract. Also, the left
inequality of $\mathtt{Contd.c}$ can be derived from the necessary
IC constraint for type $\theta_{k-1}$ in a feasible contract, and
the right inequality of $\mathtt{Contd.c}$ can also be derived from
the necessary IC constraint for type $\theta_k$. Hence,
Proposition~\ref{prop:suff} shows both the sufficient and necessary
conditions for a feasible contract. \hfill$\rule{2mm}{2mm}$



\begin{thebibliography}{1}

\bibitem{Akyildiz-survey}
I. F. Akyildiz, W-Y Lee, M. C. Vuran and S. Mohanty, ``NeXt
generation/dynamic spectrum access/cognitive radio wireless
networks: a survey,'' \emph{Computer Networks Journal (Elsevier)},
vol. 50, 2006.

\bibitem{buddhikot-sur}
M.M. Buddhikot, ``Understanding Dynamic Spectrum Access: Models,
Taxonomy and Challenges,'' in \emph{Proc. of IEEE DySPAN '07}, 2007.

\bibitem{Mahanen}
Y. Li, P. Mahanen, M. Buddhikot and Y. Liang,  ``Computer Networks
(Elsevier) Special Issue on Cognitive Wireless Networks'',
\emph{Computer Networks}, pp.775-777, 2008.


\bibitem{Marshall1}
P.F. Marshall, ``Extending the Reach of Cognitive Radio,'' in
\emph{Proceedings of the IEEE}, pp.612--625, 2009.

\bibitem{HuangLiuDing2008}
S. Huang, X. Liu and Z. Ding, ``Opportunistic Spectrum Access in Cognitive Radio Networks,'' in \emph{Proc. of IEEE INFOCOM}, 2008.


\bibitem{cr4}
Y. Xing, C.N. Mathur, M.A. Haleem, etc., ``Dynamic Spectrum Access
with QoS and Interference Temperature Constraints,'' \emph{IEEE
Trans. Mobile Computing}, vol. 6, no. 4, pp:423-433, Apr. 2007.

\bibitem{spshar1}
L. Cao and H. Zheng, ``Distributed rule-regulated spectrum
sharing,'' \emph{IEEE J. Selected Areas in Comm.}, vol. 26, no. 1,
pp. 130-145, 2008.

\bibitem{Mahonen2}
J. Riihijarvi, P. Mahonen, M. Wellens and M. Gordziel,
``Characterization and modelling of spectrum for dynamic spectrum
access with spatial statistics and random fields'', in \emph{Proc.
of PIMRC '08}, pp.1-6, 2008.

\bibitem{Mahonen3}
J. Riihijarvi,  P. Mahonen,  M.  Petrova,  V.  Kolar, ``Enhancing
cognitive radios with spatial statistics: From radio environment
maps to topology engine,'' in \emph{Proc. of IEEE CROWNCOM '09}, pp.
1-6, 2009.

\bibitem{buddhikot}
M.M. Buddhikot, ``Cognitive Radio, DSA and Self-X: Towards Next
Transformation in Cellular Networks (Extended Abstract)'', in
\emph{Proc. of IEEE DySPAN '10}, pp.1-5, 2010.

\bibitem{Pierre}
Gary De Vries and Pierre De Vries, ``The Role of Licence-Exemption
in Spectrum Reform,'' \emph{Communications \& Strategies} 67 (2007):
pp. 85-106.

\bibitem{nancy}
W. Lehr and N. Jesuale, ``Spectrum Pooling for Next Generation
Public Safety Radio Systems,'' in \emph{Proc. IEEE DySPAN '08},
pp:1-23, 2008.

\bibitem{lehr}
J.M. Chapin and W. Lehr, ``Time-limited leases for innovative
radios,'' in \emph{Proc. IEEE DySPAN '07}, pp:606-619, 2007.

\bibitem{Ballon}
S. Delaere and P. Ballon, ``Multi-level standardization and business
models for cognitive radio: the case of the Cognitive Pilot
Channel,'' in \emph{Proc. IEEE DySPAN '08}, 2008.

\bibitem{cr5}
J. Hwang and H. Yoon, ``Dynamic Spectrum Management Policy for
Cognitive Radio: An Analysis of Implementation Feasibility Issues,''
un \emph{Proc. IEEE DySPAN '08}, pp:1-9, 2008.

\bibitem{peha}
Jon M. Peha, ``Sharing Spectrum Through Spectrum Policy Reform and
Cognitive Radio,'' in \emph{Proc. of the IEEE '09}, 97 (4), pp.
708-719, 2009.


\bibitem{spectrum-trading2}
E. Hossain, D. Niyato and Z. Han, \emph{Dynamic spectrum access and
management in cognitive radio networks}, Cambridge Univ. Press, 2009.

\bibitem{Weiss}
A. Tonmukayakul and  M. B. H. Weiss, ``A study of secondary spectrum
use using agent-based computational economics,'' \emph{Economic
Research and Electronic Networking (NETNOMICS)}, 2009.

\bibitem{pricing1}
C. Kloeck, H. Jaekel, and F.K. Jondral, ``Dynamic and Local Combined
Pricing, Allocation and Billing System with Cognitive Radios,'' In
\emph{Proc. IEEE DySPAN}, 2005.


\bibitem{pricing4}
F. Wang, M. Krunz, and S. Cui,  ``Price-based spectrum management in
cognitive radio networks,'' \emph{IEEE J. Selected Topics Signal
Proc.}, 2008.

\bibitem{pricing5}
D. Niyato and E. Hossain, ``Market-equilibrium, competitive, and
cooperative pricing for spectrumsharing in cognitive radio networks:
analysis and comparison,'' \emph{IEEE Trans. Wireless Comm.}, 2008.


\bibitem{gao1}
L. Gao, Xinbing Wang, Y. Xu and Q. Zhang, ``Spectrum Trading in
Cognitive Radio Networks: A Contract-Theoretic Modeling Approach,''
\emph{IEEE J. Selected Areas in Comm.}, 2010.

\bibitem{contract2}
D.M. Kalathil and R. Jain, ``Spectrum Sharing through Contracts,''
in \emph{Proc. of IEEE DySPAN '10}, pp. 1-9, 2010.


\bibitem{JianweiHuang}
J. Huang, R. Berry and M. L. Honig,  ``Auction-based Spectrum
Sharing,'' \emph{ACM Mobile Networks and Applications Journal},
11(3), June 2006.

\bibitem{lixy}
Xiang-Yang Li, Ping Xu, ShaoJie Tang, Xiaowen Chu,``Spectrum Bidding
in Wireless Networks and Related,'' in \emph{Proc. of COCOON '08},
2008.

\bibitem{SorabhGandhi}
S. Gandhi, C. Buragohain,  L. Cao,  etc., ``A general framework for
wireless spectrum auctions,'' In \emph{Proc. of IEEE DySPAN '07},
2007.


\bibitem{gao2}
L. Gao, Y. Xu, X. Wang, ``MAP: Multi-Auctioneer Progressive Auction
in Dynamic Spectrum Access,''  \emph{IEEE Trans. Mobile Computing},
2010.

\bibitem{eBay}  X. Zhou,  S. Gandhi,  S. Suri, and H. Zheng, ``eBay in the
sky: Strategy-proof wireless spectrum auctions,'' in \emph{Proc. of
ACM MobiCom '08}, 2008.

\bibitem{TRUST} X. Zhou,  H. Zheng, ``TRUST: A General Framework for
Truthful Double Spectrum Auctions,'' in \emph{Proc. of IEEE InfoCom
'09}, 2009.

\bibitem{WangXuXu2010}
S. Wang, P. Xu, X. Xu, S.-J. Tang, X.-Y. Li and X. Liu,
``TODA: Truthful Online Double Auction for Spectrum Allocation in Wireless Networks,'' \emph{Proc. IEEE DySPAN}, 2010.

\bibitem{cocr1}
O. Simeone, I. Stanojev, S. Savazzi, Y. Bar-Ness, U. Spagnolini, R.
Pickholtz, ``Spectrum leasing to cooperating ad hoc secondary
networks,'' \emph{IEEE J. Selected Areas in Comm.}, vol. 26, no. 1,
pp. 203-213, Jan. 2008.

\bibitem{cocr2}
J. Zhang and Q. Zhang, ``Stackelberg Game for Utility-Based
Cooperative Cognitive Radio Networks,'' in \emph{Proc. ACM Mobihoc
¡¯09}, 2009.


\bibitem{cocr3}
H. Wang, L. Gao, X. Gan, Xinbing Wang, E. Hossain, ``Cooperative
Spectrum Sharing in Cognitive Radio Networks: A Game-Theoretic
Approach,'' to appear in \emph{Proc. of IEEE ICC '10}, South Africa,
2010.

\bibitem{cocr4}
Y. Han, A. Pandharipande and S.H. Ting, ``Cooperative spectrum
sharing via controlled amplify-and-forward relaying,'' in
\emph{Proc. of IEEE PIMRC '08}, pp. 1--5, 2008.

\bibitem{addition}
H. Shen and T. Basar, ``Pricing under information asymmetry for a
large population  of users,'' \emph{Telecommunication System}, May
2010.


\bibitem{Laneman}
J. N. Laneman, and G. W. Wornell, ``Distributed space-time-coded
protocols for exploiting cooperative diversity in wireless
networks,'' \emph{IEEE Trans. on Information Theory}, Oct. 2003,
49(10), 2415-2425.

\bibitem{luo2007decode}
J. Luo, R. Blum, L. Cimini, L. Greenstein and A. Haimovich,
``Decode-and-forward cooperative diversity with power allocation in
wireless networks,'' \emph{IEEE Trans. Wireless Comm.}, 2007, 6,
793-799.

\bibitem{GameTheory}
 D. Fudenberg and J. Tirole, \emph{Game Theory}, The MIT Press, 1991.

\bibitem{ContractTheory}
P. Bolton and M. Dewatripont, \emph{Contract Theory}, The MIT Press,
2005.


\bibitem{Nonlinear}
A.~Ruszczynski, \emph{Nonlinear Optimization}, Princeton Univ.
Press, 2006.



\end{thebibliography}

\end{document}